\documentclass[12pt]{article}
\usepackage{epsfig}

\parskip=\medskipamount	
\renewcommand{\thesection}{\arabic{section}}

\catcode`@=11

\@addtoreset{equation}{section}
\@addtoreset{equation}{subsection}
\def\theequation{\ifnum\value{section}=0 \arabic{equation}\ignorespaces
\else \ifnum\value{section}=-1 A.\arabic{equation}\ignorespaces
\else \ifnum\value{subsection}=0 \thesection.\arabic{equation}\ignorespaces
\else \thesection.\arabic{subsection}.\arabic{equation}\ignorespaces
                             \fi
                        \fi
                   \fi}

{\catcode`\'=\active \def'{{}^\bgroup\prim@s}}

\catcode`@=12

\newcommand{\bq}{\begin{equation}}
\newcommand{\be}{\begin{equation}} 
\newcommand{\fq}{\end{equation}}
\newcommand{\ee}{\end{equation}}
\newcommand{\bqr}{\begin{eqnarray}}
\newcommand{\beqs}{\begin{eqnarray}} 
\newcommand{\fqr}{\end{eqnarray}}
\newcommand{\eeqs}{\end{eqnarray}}

\newcommand{\rf}[1]{(\ref{#1})}

\def\bop#1{\setbox0=\hbox{$#1M$}\mkern1.5mu
	\vbox{\hrule height0pt depth.04\ht0
	\hbox{\vrule width.04\ht0 height.9\ht0 \kern.9\ht0
	\vrule width.04\ht0}\hrule height.04\ht0}\mkern1.5mu}
\def\Box{{\mathpalette\bop{}}}                        

\begin{document}
\thispagestyle{empty} 

\begin{flushright} 
\begin{tabular}{l} 
ANL-HEP-PR-01-025 \\
hep-th/0104132 \\ 
\end{tabular} 
\end{flushright}  

\vskip .3in 
\begin{center} 

{\Large\bf  M-theory and automorphic scattering} 

\vskip .3in 

{\bf Gordon Chalmers} 
\\[5mm] 
{\em Argonne National Laboratory \\ 
High Energy Physics Division \\ 
9700 South Cass Avenue \\ 
Argonne, IL  60439-4815 } \\  

{e-mail: chalmers@pcl9.hep.anl.gov}  

\vskip .5in minus .2in

{\bf Abstract}  
\end{center} 

The strongly coupled limit of string scattering and the automorphic construction 
of the graviton S-matrix is compared with the eleven dimensional formulation of 
M-theory.  In a particular scaling limit at strong string coupling, M-theory is 
described by eleven-dimensional supergravity which does not possess a dilaton, 
but rather a perturbative expansion in the gravitational coupling and derivatives.  
The latter theory provides an off-shell description of the string, upon dimensional 
reduction. 

\setcounter{page}{0} 
\newpage 
\setcounter{footnote}{0} 

\section{Introduction} 
 
M-theory has emerged as the unifying framework of the five fundamental
consistent string theories \cite{Witten:1995zh}.  At low energies,
M-theory  limits to ${\cal N}=1$ eleven dimensional supergravity
\cite{Cremmer:1978km};  more precisely  graviton scattering elements
have been computed to an order of twelve derivatives where it has been
shown to be in agreement with the superstring theory to this order
upon dimensional reduction\footnote{These calculations prohibit massive 
modes from contributing at tree-level in eleven dimensions, as would be 
expected based on general covariance.}.  ${\cal N}=1$  $d=11$ supergravity
contains the massless gravitational multiplet, and massive multiplets
have not  been consistently formulated in $d=11$ which is a strong
constraint on the matter structure (there is one  exception for a
massless matter multiplet \cite{Nishino:2000cv}.)   In addition,
string theory graviton scattering  elements must obey consistency
requirements with respect to the U-duality groups \cite{Hull:1995ys};
the strongly  coupled limit of IIA on a circle (T-dualized to IIB
superstring  theory on a circle) defines M-theory at all scales.
Given the consistency with supersymmetry and  automorphic functions, a
question is to what extent M-theory is described solely by eleven
dimensional  supergravity (an examination of some of the superspace
constraints is found in \cite{Peeters:2000sr,Gates:2001hf,Nishino:2001mb},  and
there, further corrections beyond the massless gravitational multiplet
are not permitted).  A consistent  quantum field theory description of
scattering amplitudes also generates an  off-shell description of
string theory scattering elements.

M-theory is built from the U-duality structures, and the S-matrix
which is compliant with S- and U-duality  may be expanded in the
strongly coupled regime to obtain a perturbative description of
M-theory.  M-theory is known to agree with maximal supergravity in
$d=11$  up to twelve derivatives
\cite{Russo:1997mk,Green:1997as,Tseytlin:2000sf},  as explicit
computations in supergravity and superstring theory indicate
\cite{Green:1997di,Green:1997as,Green:1997tv,Green:1998tn,Green:2000pu,Green:2000pv}.
We examine the perturbative  form of the automorphic functions in this
limit and  show compatibility with $d=11$ supergravity.   A  question
we address is how the description  differs from maximal ${\cal N}=1$
($N=32$) supergravity in eleven dimensions within a specific
first-quantized  string-inspired regulator (the form of which is very
controlled by supersymmetry constraints). The  generating functional
discussed previously is a means to  deriving dynamics at all values of
the coupling  constants in addition to providing an off-shell
description.

The automorphic string scattering reformulates the expansion at
multi-genera and packages coefficients  in the derivative expansion
(including instanton results) in terms of automorphic functions.
Taking these  results and demanding self-consistency allows for a
determination of these coefficients without resort  to the path
integral.

The outline of this work is as follows.  In section 2 we review the
automorphic basis upon which a  string S-matrix may be expanded, with
duality manifest.  In section 3 we examine the perturbative structure
of M-theory arising from the automorphic functions and demonstrate
compatibility with supergravity.  In section 4 we take the strongly
coupled limit of the T-dualized formulation of this theory, IIA to
M-theory.  We then  briefly comment on the reformulated string
expansion and sewing to obtain these coefficients.

\section{Automorphic scattering of graviton scattering}

We briefly review the construction of the S-duality compliant graviton
scattering at the four-point order in type  IIB superstring theory
(reviewed in \cite{Obers:1999fb,Obers:2000um} to eight derivatives) at
all genera  \cite{Chalmers:2000zg}.  T-dualizing along a compact
direction generates the  scattering in type IIA.  The  automorphic
construction relies upon a basis of automorphic functions that is
compatible with the perturbative  structure of the superstring (see
\cite{Terras} for a review of these functions).   A basis is
determined from the  ring of functions,  \bqr   g_k\in \prod_{j=1}^n
E_{s_j}^{(q_j,-q_j)}(\tau,\bar\tau)  \ ,
\label{modularring}
\fqr  satisfying the properties,  \bqr   \sum_{j=1}^n q_j = 0 \qquad
\sum_{j=1}^n s_j = s \ ,   \fqr  and invariance under  \bqr
\tau\rightarrow {a+b\tau\over c+d\tau} \ ,  \fqr  with $bc-ad=1$.
The first property enforces modular invariance, and $q_i$ is related
to the $U(1)_R$ charge.    The second parameterizes the asymptotic
structure.  The ring of functions is specified by partitioning  an
integer $s$ into half-integers with $\vert q\vert\leq s$, and the
functions themselves are algebraic,  \bqr
E_s^{(q,-q)}(\tau,\bar\tau) = \sum_{(m,n)\neq (0,0)} {\tau_2^s\over
(m+n\tau)^{s+q} (m+n\bar\tau)^{s-q}} \ .
\label{eisenstein}
\fqr  The asymptotics of these functions are,  \bqr
E_s^{(q,-q)}(\tau,\bar\tau) = g_0 \tau_2^s + g_1 \tau_2^{1-s} + {\cal
O}(e^{-2\pi \tau_2}) \ ,
\label{esexpand}
\fqr  where the coefficients in \rf{esexpand} are,   \bqr   g_0 = 2
\zeta(2s) \qquad\qquad g_1 = 2\sqrt{\pi} \zeta(2s-1)
{\Gamma(s-{1\over 2})\over \Gamma(s)} \quad, \qquad \zeta(s) =
\sum_{n=1}^\infty {1\over n^s} \ .   \fqr  and describe the
perturbative terms together with exponentially decaying instanton (and
wrapped solitons  on compact spaces) corrections to graviton
scattering.  We restrict to the cases of half-integral or integral
$s_j$ and do not include the cusp forms; cusp functions have only
instantonic or exponential terms in the  asymptotic expansion.  The
functions in the ring in \rf{modularring} have the asymptotics,  \bqr
g_k = g^{(0)}_k \tau_2^{{3\over 2}+{k\over 2}} + g^{(1)}_k
\tau_2^{-{1\over 2}+{k\over 2}} + \ldots +  g^{(h)}_k \tau_2^{{3\over
2}+{k\over 2}-2h} + {\cal O}(e^{-2\pi\tau_2})  \ ,  \fqr  where $h$ is
correlated with a maximum genus on the string S-matrix expansion.
S-duality and a direct  convergence of the dual planar limit of ${\cal
N}=4$ super Yang-Mills theory defined through covariantized
holographic string scattering \cite{Chalmers:2000vq} in the AdS/CFT
correspondence  \cite{Maldacena:1998re,Gubser:1998bc,Witten:1998qj}
support that there are no further  coefficients beyond the maximum
genus $h$, $h={1\over 2}(k+2)$ for $k$ even and ${1\over 2}(k+1)$ for
$k$ odd;  further modular forms in the complete basis,
\rf{eisenstein}, composed from $s$ on the complex plane do not agree
in general with the perturbative structure of the string.

These modular invariant functions may be generalized from the
fundamental domain  $U(1)\backslash SL(2,R)/SL(2,Z)$ describing the
vacuum of uncompactified type IIB superstring  theory to  toroidally
compactified spaces, and pertain to other examples, including type IIB
propagating on K3\footnote{The  metric on K3 is not known, but the
moduli space of manifolds is known, being a simple coset, and the
latter is  sufficient to defing automorphic functions on a K3.  This
potentially allows for a reconstruction of the K3 metric  from the
scattering.}  The  graviton scattering S-matrix must be expanded on
these functions.  At the four-point  order, on-shell supersymmetry
restricts the tensor structure to be composed of the Bel-Robinson form
\cite{BelRobinson},  with the tensor $R^4$ built out of contracting
eight Weyl tensors as in \cite{Green:1987sp},   \bqr   S= \int
d^{10}x\sqrt{g} ~\Bigl[ {1\over\Box^3} R^4 + \sum_{n=0}^\infty
g_k(\tau,\bar\tau)  \left(\alpha'\Box\right)^n R^4 \Bigr] \ ,
\label{local}
\fqr  where the tensor structure of the boxes is left undetermined.
At tree-level the derivative structure has the  form,  \bqr
s^k+t^k+u^k \ ,   \fqr  but in general there are mixed products of the
invariants,  \bqr  s^{k_1}t^{k_2}u^{k_3} \qquad k=k_1+k_2+k_3 \ .
\fqr  The $R^4$ tensor has the form,  \bqr  t_8^{\mu_1\ldots\mu_8}
t_8^{\nu_1\ldots\nu_8} k_{1,\mu_1\nu_1} k_{2,\mu_3\nu_3}
k_{3,\mu_5\nu_5} k_{4,\mu_7\nu_7}
\prod_{h=1}^4\varepsilon_{h,\mu_{2h}\nu_{2h}} \ ,  \fqr  with  \bqr
t_8^{\mu_1\ldots\mu_8}= -{1\over 2}\epsilon^{\mu_1\ldots\mu_8}  \fqr
\bqr  -{1\over 2} \Bigl\{ \left(\eta^{\mu_1\mu_3} \eta^{\mu_2\mu_4}
-\eta^{\mu_1\mu_4}\eta^{\mu_2\mu_3}\right)   \left(\eta^{\mu_5\mu_7}
\eta^{\mu_6\mu_8} -\eta^{\mu_5\mu_6}\eta^{\mu_7\mu_8}\right)  \fqr
\bqr  + \left(\eta^{\mu_3\mu_5}\eta^{\mu_4\mu_6} -
\eta^{\mu_3\mu_6}\eta^{\mu_4\mu_5} \right)
\left(\eta^{\mu_7\mu_1}\eta^{\mu_8\mu_2}-\eta^{\mu_7\mu_2}\eta^{\mu_1\mu_8}\right)
\fqr  \bqr  + \left( \eta^{\mu_1\mu_5}\eta^{\mu_2\mu_6} -
\eta^{\mu_1\mu_6}\eta^{\mu_2\mu_5}\right)
\left(\eta^{\mu_3\mu_7}\eta^{\mu_4\mu_8} -
\eta^{\mu_3\mu_8}\eta^{\mu_4\mu_7}\right)   \fqr  \bqr
\hskip -.4in  + {1\over 2} \Bigl(
\eta^{\mu_2\mu_3}\eta^{\mu_4\mu_5}\eta^{\mu_6\mu_7}\eta^{\mu_8\mu_1} +
\eta^{\mu_2\mu_4}\eta^{\mu_6\mu_3}\eta^{\mu_4\mu_7}\eta^{\mu_8\mu_1} +
\eta^{\mu_2\mu_4}\eta^{\mu_6\mu_7}\eta^{\mu_3\mu_8}\eta^{\mu_4\mu_1}
\Bigr) + {\rm perms}\Bigr\}  \fqr  There are non-analytic terms that
may be  constructed by unitarity from the local terms in \rf{local}
\cite{Chalmers:2000zg},  and the form in \rf{local} has manifest
duality invariance under the fractional linear transformations when
the energy scales are below the regime $s_{ij} \leq 4/\alpha'$.  The
massive modes appear upon resumming  the higher derivative terms, as
in the expansion of $\ln(1-\alpha's)$.

The generalization to the $SL(d,Z)$ modular forms is parameterized by
the moduli of a d-dimensional torus  and in the limit of one of the
radii becoming large we have the expansion \cite{Obers:1999fb,Obers:2000um},  
\bqr   
E^{\rm SL(d,Z)}_{R=d,s}(\phi_j) = 
E^{\rm SL(d-1,Z)}_{R=d-1,s}(\tilde\phi_j) +
{2\pi^s\Gamma(s-{d-1\over 2}) \zeta(2s-d+1) \over \pi^{s-{d-1\over 2}}
\Gamma(s) R^{2s-d+1} V_{d-1} }   \fqr  \bqr    + {2\pi^s\over
\Gamma(s) R^{2s-d-1}} \sum_{m^a,n^b} \vert { n^a g_{ab} n^b \over
m^2} \vert K_{s-{d-1\over 2}}\Bigl(  2\pi \vert m\vert R \sqrt{n^a
g_{ab} n^b}  \Bigr) \ .
\label{asymptSLd}
\fqr  
In the case of $SL(2,Z)$ the expansion relevant for the M-theory
limit from IIB superstring theory  compactified on a circle of radius
$R$ is,    
\bqr   E^{\rm SL(2,Z)}_{R=2,s}(\phi_j) =
{2\pi^s\Gamma(s-{d-1\over 2}) \zeta(2s-d+1) \over \pi^{s-{d-1\over 2}}
\Gamma(s) R^{2s-d+1} V_{d-1} }   \fqr  \bqr   + {2\pi^s\over \Gamma(s)
R^{2s-d-1}} \sum_{m^a,n^b} \vert { n^a g_{ab} n^b \over  m^2} \vert
K_{s-{d-1\over 2}}\Bigl(  2\pi \vert m\vert R \sqrt{n^a g_{ab} n^b}
\Bigr) \ ,
\label{asymptots}
\fqr  with $g_{ab}=\pmatrix{ 1 & 0 \cr 0 & \tau }$.   The first term
is relevant for the eleven-dimensional limit and the latter terms
decouple.

\section{Eleven dimensional quantum limit}

The IIB superstring compactified on a circle and T-dualized describes
the M-theory  limit upon taking the compact volume to infinity.  The
coupling constants are identified as  \bqr   V=R_{10} R_{11}
\qquad\quad \tau=C^{(0)}+i {R_{10}\over R_{11}} = {\theta\over 2\pi} +
i{4\pi\over g_s^2} \ ,    \fqr  \bqr   R_{11} = e^{2\phi^A/3} \qquad
R_{10} = r_A e^{\phi^A/3} \ ,  \fqr  and  \bqr   l_{11} =
(g^A)^{1\over 3} l_{10} \ ,  \fqr  where $r_A$ is a dimensionless
radii of the tenth dimension as measured in Planck units.
Furthermore, the kinematic invariants in the ten-dimensional Einstein
frame of the string and the eleven  dimensional supergravity ones are,
\bqr   s_{ij} = S_{ij} { l_{11}^2\over l_{10}^2 R_{11}} \ ,
\label{scalingMandelstam}
\fqr  in the string frame  with the inverse power of the eleven
dimensional radius (at finite values of the compact eleven dimensional
radius) arising from the inverse metric in $S_{ij} = - G^{\mu\nu}
(k_i+k_j)_\mu (k_i+k_j)_\nu$.   It will be  useful for us to S-dualize
the torus so that the coupling has the inverse $\tau_2\rightarrow
1/\tau_2$.  Then a term generated at large volume from the automorphic
functions has the coupling  dependence from,  \bqr   E_s = 2\zeta(2s)
\bigl({R_{11}\over R_{10}}\bigr)^{-s} + 2\sqrt{\pi} \zeta(2s-1)
{\Gamma(s-{1\over 2})\over \Gamma(s)} \bigl({R_{11}\over
R_{10}}\bigr)^{s-1} + {\cal O}(e^{-2\pi\tau_2}) \ ,    \fqr  \bqr
g_k(\tau,\bar\tau) l_{10}^{2k-1} \Box^k R^4 \rightarrow l_{10}^{2k-1}
\Box^k R^4  =  l_{10}^{2k-1} \bigl({R_{11}\over
R_{10}}\bigr)^{-3-k+2g_{\rm max}}   \Box^k_{10} R^4_{10}  \ ,
\label{couplings}
\fqr  where a transformation to string frame from Einstein frame is
taken and $g_{\rm max}={1\over 2}(k+2)$ for  g even and ${1\over
2}(k+1)$ for g odd.  There are subleading terms in \rf{couplings} and
a factor of  $\sqrt{\tau_2}$ is also absorbed into the latter  limit
from the determinant of the metric.  We next  examine the coupling
structure  in the loop expansion of supergravity.

Translating to loop counting via the Einstein-Hilbert term, ${1\over
l_{11}^9} \int d^{11}x \sqrt{g} R$,  generates the power series from 
the above ten-dimensional derivative terms,  \bqr   l_{11}^{2k-1+5L} 
g_A^{-{2\over 3}k -{1\over 3}}
\bigl({R_{11}\over R_{10}}\bigr)^{-3-k+2g_{\rm max}} \Box^k_{11}
R^4_{11} l_{11}^{-5L} \ ,
\label{maximalsugra} 
\fqr  where a factor of $l_{11}^{-5L}$ is taken out of the prefactor.
This latter factor nullifies the primitive  divergence in a $\phi^3$
graph in eleven dimensions; the divergence structure of maximal
supergravity is conjectured  to be that of $\phi^3$ theory with
$2(L-1)$ kinematic invariant factors in front when regulated in a
string inspired regulator (e.g. $R^4$ at one-loop, $s^2 R^4$  at two
loops, etc...) \cite{Bern:1998ug,Chalmers:2000ks}.  The
string-inspired regulator  is elucidated in \cite{Green:2000pu} at
two-loops.   The divergence structure in eleven dimensions of maximal
supergravity then agrees  with the couplings in \rf{maximalsugra} if
$k=2L$, as the coupling constant in maximal  supergravity in eleven
dimensions is $\kappa_{11}^2 = l_{11}^9$.  Thus the maximal loop graph
contributing according  to the M-theory limit is in agreement with the
conjectured form of the supergravity expansion, after including the 
additional eleventh dimensional volume element.  (The additional
factor of $l_{11}^{-1}$ is a remnant of taking the limit from nine
dimensions and two additional powers  appear in formulating in eleven
dimensions.)

Example maximum couplings up to order $k=4$ are,  \bqr   {1\over
\Box^3} R^4 \qquad : \qquad L=0  \ ,  \nonumber  \fqr  \bqr
E_{3\over 2}~ R^4 \qquad : \qquad L=1 \ ,  \nonumber  \fqr  \bqr
E_{5\over 2} ~\Box^2 R^4 \qquad\quad :  L=2 \ ,  \nonumber  \fqr  \bqr
E_3 \Box^3 R^4, \vert E_{3/2}^{(1,-1)}\vert^2 \Box^3 R^4 \qquad : L=2
\ ,  \fqr  \bqr   \ldots \quad :\quad ~\ldots \ ,  \fqr   and at
higher powers the terms are descending power series in multiples of
four.  The genus one and  two results for the $E_{5/2}$ function have
been examined in
\cite{Green:1997tv,Green:1998tn,Green:2000pu,Green:2000pv}.   The
$1/\Box^3 R^4$ is two derivatives and arises at tree-level in the
graviton scattering.  The string  perturbative expansion is in powers
of $\tau_2^2$.   This table is in agreement with the known
supergravity  structure in $d=11$  dimensions.   The coefficients at
arbitrary genera are not known but the cancellation  properties of the
expansion  are in agreement with this power series and the instanton
corrections.

Matrix theory \cite{Banks:1997vh} is a partonic description of the
D-brane in M-theory, defined by taking the  dimensional reduction of
ten-dimensional supergravity, in the same vein as Nahm's equations
are the dimensional reduction of gauge theory and describe solitons.  
The automorphic construction, and the quantum regulated supergravity, 
agrees with the eleven dimensional origin of Matrix theory in the large 
N limit.  Here the 
analogy is between four-dimensional gauge theory and eleven dimensional 
supergravity and the states of the dyons or solitons in the gauge theory 
or string theory described by the dimensional reduced theory to one 
dimension.

\section{Recursive construction of the amplitudes} 

The automorphic properties of the scattering and a recursive
formulation without a  direct resort to the functional integral also
obeys this correspondence with the $N=1$ $d=11$ supergravity.  In this
section we reformulate the expansion and derive a recursive algorithm
for the coefficients.    We expand on this in future work
\cite{inprogress}.

The basic identity for recursively deriving the coefficients follows
from sewing the S-matrix onto  itself: $\sum SS=S$ with a summation
over the intermediate lines both in number and particle type.  This
sewing  relation involves only one-loop integrals in order to obtain
multi-genus string theory results; the  string loop integrals have
been performed and encoded in the automorphic function basis,
extracting  the manifest non-perturbative duality and symmetries of
the string as well as the instanton coefficients  related to the genus
expansion via modular invariance.  The primary complication is the
tensor contractions  associated with the massive string modes,
however, string field theory techniques may be adapted to  generate
this sewing.

The generating functional for $\Phi^3$ theory is an example that does
not have the complications  with the tensors.  We integrate out the
loops and perform a derivative expansion, obtaining the  on-shell
amplitudes,  \bqr   A_4(x_j) = \sum_{k=2}^\infty \int A_{2+k}(x_1,x_2;y_j)
~\prod_{j=1}^k \Delta(y_j-z_j) ~ A_{2+k}(z_j;x_3,x_4) \vert_{y_j=y,z_j=z} 
\label{sewingphicubed} 
\fqr  power series expanded in derivatives from,  \bqr   A_n =
\sum_{a,b} ~ c_{a,n}^b t_{a,n}^b \ ,
\label{expanded}
\fqr  with the invariants,  \bqr  t_{i,m} = \left( k_i+ \ldots +
k_{i+m-1}\right)^2  \qquad \left(t_{i,2}=s_{i,i+1)} \right) \fqr (the series 
in \rf{expanded} is a truncated one and not the most general).  The graphs 
in \rf{sewingphicubed} is the s-channel recursion. 
Inserting the derivative expansion of the amplitude into the sewing
relation \rf{sewingphicubed} gives  recursive relations relating loops
at different orders ($l_1+l_2=l_3$ with $l_1$ and $l_2$ the loop
orders of the two amplitudes),  \bqr  \sum_{a,b} c_{a;4}^b t_{a,4}^b =
\sum_{i,l,{\tilde l},m} \int c_{i,m} \Bigl( \sum_{n=1}^{m-1} \partial_{i+n}^y
\Bigr)^{2l} \prod_{j=1}^k \Delta(y_j-z_j)  \sum_{{\tilde l},{\tilde
i}} c_{{\tilde i},m} \Bigl( \sum_{n=1}^{m-1} \partial_{{\tilde i}+n}^z
\Bigr)^{2{\tilde l}} \vert_{y_j=y,z_j=z} \ ,
\label{expandedsewingphicubed} 
\fqr  
integrated with $d^4(y-z)$ and the directions of the derivatives implied on 
the propagators and the uninserted external lines.  
The derivatives in \rf{expandedsewingphicubed} and their tensor
form are written in compact form because  of the many propagators.  The 
$k$ massive propagators runs up to $b-2{\tilde l}-2l-d$, with $d$ the 
dimension of the integrated $d^4(y-z)$ spacetime.  The expression is 
more familiar in k-space, however, in x-space the integrals are more 
explicit; An interesting feature is that in $k$-space the integrals are 
multi-loops and  complicated to evaluate but in $x$-space all of the 
integrals are {\it one-loop} and may be performed explicitly.  The coefficients 
may  be iteratively constructed from
one loop order to the next.  This is the sewing relation we generalize
now to the context of the superstring, which lifts directly into eleven 
dimensions, together with manifest duality
properties encoded in the automorphic  functions.  In the superstring
this formalism also avoids complications with superconformal ghost
systems;   however, superconformal insertions are potentially useful
in mechanizing cancellations in supergravity  from the superstring
\cite{Chalmers:2000ks}.

The graviton scattering from the covariantized S-matrix in \rf{local}
is generated by the derivatives,  \bqr   A_4(k_i,\epsilon_i) =
\prod_{j=1}^4 \varepsilon^{\mu_j\nu_j}   {\partial\over\partial{\hat
g}^{\mu_j\nu_j}} S[{\hat g}+\eta] \vert_{{\hat g}=0} \ ,
\label{fourpoint} 
\fqr  after including the non-analytic terms derived from iterating
unitarity cuts in \rf{local} \cite{Chalmers:2000zg}.  The  four-point
amplitude in \rf{fourpoint} holds at arbitrary genera and in order to
formulate the sewing  for the generating functional we require the
similarly formulated amplitudes with massive external string  states
and higher-point amplitudes.

As an example, the four-point classical Veneziano-Shapiro amplitude
\cite{Green:1987sp} in the limit  $\alpha'\rightarrow 0$ is, in string
frame,   \bqr   A_4 = e^{-2\phi} R^4
{\Gamma(1+\alpha's)\Gamma(1+\alpha't)\Gamma(1+\alpha'u)\over
\Gamma(1-\alpha's) \Gamma(1-\alpha't)\Gamma(1-\alpha'u)} = e^{-2\phi}
R^4 + \ldots \ ,
\label{VSamplitude} 
\fqr  with a derivative expansion from the covariantized scattering
amplitude,  \bqr   A_{4,g=0}^{\rm IIB} = 64 e^{-2\phi} {R^4\over
\alpha'^3 stu} {\rm exp}\Bigl(  \sum_{p=1}^\infty {2\zeta(2p+1)\over
2p+1} \left({\alpha'\over 4}\right)^{2p+1}  \left( s^{2p+1} + t^{2p+1}
+ u^{2p+1} \right) \Bigr) \ .
\label{treeexpform}  
\fqr  The first term in \rf{VSamplitude} is deduced from sewing the
three- and four-point variations of $\int d^{10}x  \sqrt{g}~ R$.  The
expansion of the \rf{VSamplitude} gives contributions at all
derivative orders in the expansion  of \rf{fourpoint}.

An individual term in the expansion is, with $g=\eta+{\hat g}$,  \bqr
V_{ \{ \mu_j\nu_j\} } = \prod_{j=1}^4
{\partial\over\partial\eta^{\mu_j\nu_j}} \sqrt{g} \Box^m R^4  \ ,
\fqr  with a modular function prefactor.   The higher point vertices
we define in the following with string field techniques.  Similar
construction  of the vertices can be found from the Koba-Nielsen
formula.  The propagators in de Donder gauge are  \bqr
\Delta^{\mu_1\nu_1,\mu_2\nu_2}(x_1,x_2) = -{1\over 2}{1\over
(x_1-x_2)^2} \left( \partial_\mu g_{\nu\sigma}  + \partial_\nu
g_{\mu\sigma} - \partial_\sigma g_{\mu\nu} \right)  \fqr  The
contribution to the four-point function is  \bqr  A_4=
\tau_2^{{3\over 2}+{k\over 2}-2g_1} \tau_2^{{3\over 2}+{k\over
2}-2g_2} ~ \varepsilon^{\mu_1\nu_1}\varepsilon^{\mu_2\nu_2}
\varepsilon^{\alpha_3\beta_3} \varepsilon^{\alpha_4\beta_4}
k_1^{\mu_5} k_1^{\nu_5} k_2^{\mu_6} k_2^{\nu_6}  k_3^{\alpha_5}
k_3^{\beta_5} k_4^{\alpha_6} k_4^{\beta_6}  \nonumber  \fqr  \bqr
\times  \int d^{10}x  V_{\mu_1\nu_1,\mu_2\nu_2;\mu_3\nu_3,\mu_4\nu_4}
\vert_{x_1} ~\Delta^{\mu_3\alpha_3,\nu_3\beta_3}
\Delta^{\mu_4\alpha_4;\nu_4\beta_4}~
V_{\alpha_1\beta_1,\alpha_2\beta_2;\alpha_3\beta_3,\alpha_4\beta_4}
\vert_{x_2} \ ,  \fqr  with $x=x_1-x_2$.  From the genus $g_1$ and
$g_2$ results including the supersymmetric intermediate states, we may
obtain a  contribution to the genus $g_1+g_2$ graviton scattering
contribution.   \bqr   A_4 =  ~ t_{8,\mu_1\ldots\mu_8}
t_{8,\nu_1\ldots\nu_8} k_1^{\mu_5} k_1^{\nu_5} k_2^{\mu_6} k_2^{\nu_6}
k_3^{\mu_7} k_3^{\nu_7} k_4^{\mu_8} k_4^{\nu_8} I(k_j) \ .  \fqr  The
inclusion of the massive multiplets allows for the derivation of
arbitrary genus results.  The sewing relations of the string 
scattering lift into the eleven-dimensional limit within the derivative 
expansion (where the massive string modes, or more generally branes 
on cycles, decouple).  

\section{Discussion} 

We have examined the eleven-dimensional quantum limit from the
automorphic construction of the IIB superstring  S-matrix elements.
The limit generates a route to a description of M-theory in terms of
perturbative eleven dimensional  supergravity.  Upon toroidal
compactification the automorphic S-matrix elements map to string
theory ones, as verified  explicitly to twelve derivatives, and
permits an off-shell description of the string scattering through a
quantum  field theory description.  Further calculations at
three-loops in supergravity or at genus two are required to  test the
duality equivalence.  The coefficients in the expansion of the
Eisenstein functions are generated from the  massive and massless
multiplets in the string, but the coefficients are in correspondence
with the quantum eleven  dimensional massless limit.  The recursive
construction of the amplitudes following from sewing the derivative
expanded form of the S-matrix generates multi-genus results from field
theory ones \cite{inprogress} and is examined  briefly in this work.

\section*{Acknowledgements} 

The work of GC is supported in part by the US Department of Energy, Division of High 
Energy Physics, contract W-31-109-ENG-38.


\begin{thebibliography}{99} 

\bibitem{Witten:1995zh}
E.~Witten,
hep-th/9507121.

\bibitem{Cremmer:1978km}
E.~Cremmer, B.~Julia and J.~Scherk,
Phys.\ Lett.\ B {\bf 76}, 409 (1978).

\bibitem{Nishino:2000cv}
H.~Nishino,
Phys.\ Lett.\ B {\bf 492}, 201 (2000)
[hep-th/0008029].

\bibitem{Hull:1995ys}
C.~M.~Hull and P.~K.~Townsend,
Nucl.\ Phys.\ B {\bf 438}, 109 (1995)
[hep-th/9410167].

\bibitem{Peeters:2000sr}
K.~Peeters, P.~Vanhove and A.~Westerberg,
hep-th/0010182.

\bibitem{Gates:2001hf}
S.~J.~Gates and H.~Nishino,
hep-th/0101037.

\bibitem{Nishino:2001mb}
H.~Nishino and S.~Rajpoot,
hep-th/0103224; 
M.~Cederwall, U.~Gran, M.~Nielsen and B.~E.~Nilsson,
JHEP {\bf 0010}, 041 (2000)
[hep-th/0007035].

\bibitem{Russo:1997mk}
J.~G.~Russo and A.~A.~Tseytlin,
Nucl.\ Phys.\ B {\bf 508}, 245 (1997)
[hep-th/9707134].

\bibitem{Green:1997as}
M.~B.~Green, M.~Gutperle and P.~Vanhove,
Phys.\ Lett.\ B {\bf 409}, 177 (1997)
[hep-th/9706175].

\bibitem{Tseytlin:2000sf}
A.~A.~Tseytlin,
Nucl.\ Phys.\ B {\bf 584}, 233 (2000)
[hep-th/0005072].

\bibitem{Green:1997di}
M.~B.~Green and P.~Vanhove,
Phys.\ Lett.\ B {\bf 408}, 122 (1997)
[hep-th/9704145].

\bibitem{Green:1997tv}
M.~B.~Green and M.~Gutperle,
Nucl.\ Phys.\ B {\bf 498}, 195 (1997)
[hep-th/9701093].

\bibitem{Green:1998tn}
M.~B.~Green and M.~Gutperle,
JHEP{\bf 9801}, 005 (1998)
[hep-th/9711107].

\bibitem{Green:2000pu}
M.~B.~Green, H.~Kwon and P.~Vanhove,
Phys.\ Rev.\ D {\bf 61}, 104010 (2000)
[hep-th/9910055].

\bibitem{Green:2000pv}
M.~B.~Green and P.~Vanhove,
Phys.\ Rev.\ D {\bf 61}, 104011 (2000)
[hep-th/9910056].

\bibitem{Obers:1999fb}
N.~A.~Obers and B.~Pioline,
Phys.\ Rept.\ {\bf 318}, 113 (1999)
[hep-th/9809039].

\bibitem{Obers:2000um}
N.~A.~Obers and B.~Pioline,
Commun.\ Math.\ Phys.\ {\bf 209}, 275 (2000)
[hep-th/9903113].

\bibitem{Chalmers:2000zg}
G.~Chalmers,
Nucl.\ Phys.\ B {\bf 580}, 193 (2000)
[hep-th/0001190].

\bibitem{Terras}  
A.\ Terras, Harmonic analyis on symmetric spaces and applications, I and II, 
Springer-Verlag, 1985.     

\bibitem{Chalmers:2000vq}
G.~Chalmers and J.~Erdmenger,
Nucl.\ Phys.\ B {\bf 585}, 517 (2000)
[hep-th/0005192].

\bibitem{Maldacena:1998re}
J.~Maldacena,
Adv.\ Theor.\ Math.\ Phys.\  {\bf 2}, 231 (1998)
[hep-th/9711200].

\bibitem{Gubser:1998bc}
S.~S.~Gubser, I.~R.~Klebanov and A.~M.~Polyakov,
Phys.\ Lett.\  {\bf B428}, 105 (1998)
[hep-th/9802109].

\bibitem{Witten:1998qj}
E.~Witten,
Adv.\ Theor.\ Math.\ Phys.\  {\bf 2}, 253 (1998)
[hep-th/9802150].

\bibitem{BelRobinson}
L.\ Bel, Acad. Sci. Paris, Comptes Rend. {\bf 247}:1094 (1958), 
{\bf 248}:1297 (1959); I.\ Robinson, unpublished.  

\bibitem{Green:1987sp}
M.~B.~Green, J.~H.~Schwarz and E.~Witten,
``Superstring Theory. Vol. 1: Introduction,'';
``Superstring Theory. Vol. 2: Loop Amplitudes, Anomalies And Phenomenology,''
{\it  Cambridge, Uk: Univ. Pr. ( 1987) 469, 596 P. ( Cambridge Monographs On Mathematical Physics)}. 

\bibitem{Bern:1998ug}
Z.~Bern, L.~Dixon, D.~C.~Dunbar, M.~Perelstein and J.~S.~Rozowsky,
Nucl.\ Phys.\  {\bf B530}, 401 (1998)
[hep-th/9802162].

\bibitem{Chalmers:2000ks}
G.~Chalmers,
hep-th/0008162.

\bibitem{Banks:1997vh}
T.~Banks, W.~Fischler, S.~H.~Shenker and L.~Susskind,
Phys.\ Rev.\ D {\bf 55}, 5112 (1997)
[hep-th/9610043].

\bibitem{inprogress} 
G.~Chalmers, in progress.

\end{thebibliography}
\end{document}